Cluster-Variational  Treatment  of Disordered  Mixed Spin Ising Model

Sobhendu K. Ghatak

Department of Physics & Meteorology

Indian Institute of Technology, Kharagpur 721302 ,India

ABSTRACT

A disordered alloy $A_p B_{1-p}$  where both A and B represent the magnetic atoms with respective spin $S_A =1/2$ and  $S_B =1$  and  whose  magnetic  interaction  can  be  described  through  Ising Hamiltonian is treated using the cluster-variational method. In  this method it is assumed that the system is built out of " building block" which is embedded in an effective field. Taking " building block"  as 4-atom cluster the approximate free energy of the alloy is then obtained by treating  the  interactions between spins within the cluster of all possible configurations in exact manner and the rest of the interaction by an effective variational field . The magnetization   M and  transition  temperature  $Tc$  are  then  calculated  for  different  concentration  and  exchange parameters ($J_{AA}$, $J_{BB}$ and $J_{AB}$ ). The magnetization   M exhibits different kinds of ferrimagnetic behaviour depending on concentration and  relative strength of intra- and inter- sub-network exchange interactions. For antiferromagnetic $J_{AB}$ ,the sub-network magnetization saturates and aligned antiferromagnetically at low temperature. The existence of compensation temperature $T_{cm}$, where total magnetization reverses its direction, depends sensitively on relative values  of $J_{AA}$/ $J_{AB}$ and   $J_{BB}$/ $J_{AB}$ and p. For B (A)-rich alloy with small $J_{AB}$, the direction of net magnetization remains same upto $T_c$ and a maximum of M  appears at intermediate   $T < T_c$ when $J_{BB} >> J_{AA}$ ($J_{BB} << J_{AA}$ ).W  hen $\left| J_{AB} \right| > J_{AA}$ , $J_{BB}$, $T_c$ exhibits  a maximum with p. The transition temperature is much less than the mean-field value for all cases. The magnetic susceptibility for $T \geq T_c$  diverges and is Curie-Wiess like at $T >> T_c$. The meta-magnetic behaviour at  high magnetic field  has been found. Some of these results are   in tune with experimental observation in amorphous rare-earth-transition metal alloys.

PACS : 05.50. +9;75.10Hk;75.10.-b

**Keywords**: Mixed Spin Ising model, Disordered ferrimagnetic alloy, Rare-earth-transition metal alloy, Cluster-variational method

## 1.INTRODUCTION

Ferrimagnetic state in simple form is characterized to a by opposing and unequal magnetization of two sublattices, and the net magnetization exists below a critical temperature. Difference in magnetic moment of constituents metal ions of the material mainly leads to unequal magnetization. In addition, differences in rate of thermal demagnetization of sublattice magnetization can result complete cancellation at lower temperature –referred as compensation temperature. The compensation point has been observed in number of ferromagnetic system is of interest from application point of view[1]. Ferrimagnetic insulator to a first approximation can.be modeled as mixed spin system $A_p B_{1-p}$ with two sublattice network. In crystalline phase the magnetic ions A with spin $S_A$ and B with spin $S_B$ occupy their respective sublattice sites. On the other hand in disordered state (like amorphous state) of a binary alloy the site occupancy tends to become random in nature. With the development of rapidly quenched technique random magnetic alloy have been intensively studied [2,3]. The composition of alloy can be varied over a wide range in amorphous state produced through rapidly quenched method. The real alloy contains, apart from magnetic atom, glass former that stabilizes the disordered state. The amorphous alloy with two kinds of magnetic atom can be, to a first approximation, considered as binary alloy as glass former is non-magnetic in nature. One of the interesting system of such alloy is rare-earth –transition metal [4-6] - like Gd-Fe, Er-Fe alloy. Rare-earth and Fe- ions carry different magnetic moment. Due to antiferromagnetic interaction between the rare earth and transition metal these alloy exhibits ferrimagnetic behaviour [4-6]. With increasing temperature the magnetization of Er subnetwork falls off rapidly compared to that of transition metal (Fe) and this results a complete compensation of magnetization of two subnetworks at a compensation temperature $T_{cm}$. It is found that for Er-Fe alloy the $T_{cm}$ decreases with decrease in concentration of Er and for larger concentration the reversal of magnetization does not occur [7].. In a-$Gd_{1-p}$-$Fe_p$ system the transition temperature $T_c$ exhibit a broad maximum for $0.2 < p < 0.6$ [8]. The total magnetization at low temperature is dominated by Gd subnetwork and decreases to zero at $T_{cm}$ and between $T_{cm}$ and $T_c$ transition metal magnetization determines the magnetic state [9]. The experimental results on these systems are obtained for limited range of concentration [8,9].

Theoretical model frequently utilized to describe the global phase diagram and the magnetic behaviour of disordered magnetic alloy is disordered Ising model [10-14]. Recently the mixed spin Ising model is being studied for a simpler model for ferrimagnetism. Different theoretical methods like mean-field approximation [15] effective field theories [16-17], renormalization-group calculations [18-19] and Monte-Carlo simulations [20,21] are utilized to get the phase diagram and critical behaviour of Ising model with spin $S = 1/2$ and $S = 1$ for two sublattices in ordered lattice. The model has also been studied in different decorated lattices [22,23] and it is predicted that the compensation temperature exists within a specific region of $J_{AA} - J_{BB}$ plane where $J_{AA}$ $(J_{BB})$ is intra-sublattice exchange interaction in A- (B) sublattice[23].

In this article the magnetic behaviour of mixed spin Ising model that can be associated with binary magnetic alloy $A_p B_{1-p}$ where the magnetic atoms A and B carry different spin $S_A$ and $S_B$. We calculate the free energy using the approximate procedure proposed by Oguchi [24] and extended for disordered system by Ghatak[14] . The magnetic properties like magnetization ,transition and compensation temperatures, susceptibility etc. are then obtained from the configuration averaged free energy. Depending upon the concentration p and relative strength of intra- ( $J_{AB}$) and intra- ( $J_{AA}$ , $J_{BB}$) exchange interactions different types of ferromagnetic behaviour have been found. The compensation temperature exits in limited region of phase space defined by exchange interaction and depends on external magnetic field. A formula for susceptibility is obtained from numerical results for $T \geq T_c$. In sec.2 the model and the method of calculation are outlined and results are presented in Sec.3.

## 2. MODEL AND METHOD OF CALCULATION

We take a binary alloy $A_p B_q$ of two magnetic atoms A and B with respective concentration p and q =1-p. It is assumed that all magnetic interactions are localized and can be described by Ising Hamiltonian

$$H = -\sum_{ij} \ J_{ij} \ S_{iz} \ S_{jz} \ - \sum_i H_i \ S_{iz} \qquad (1)$$

Where $S_{iz}$ is the Ising spin at i-th site and takes the value $S_A$ or $S_B$ depending upon the occupation of the site by A or B atom. As the mixed spin system is characterized by different magnetic moment of the constituent we simulate this situation assuming $S_A = \frac{1}{2}$ and $S_B = 1$. These values are chosen to reduce the algebraic complexity. The second term is the Zeeman energy where the magnetic field Hi (expressed in dimension of energy) at i-th site and represents sum of anisotropic field and external field $H_0$ in z-direction. The nearest neighbour exchange interaction $J_{ij}$ takes value $J_{AA}$, $J_{BB}$ and $J_{AB}$ for the magnetic bond A-A, B-B and A-B respectively. Assuming quenched disorder the thermodynamics of the disordered alloy can be obtained from the free energy F, given by [14]

$$F = - \ kT \ [ \ ln \ Tr \ exp \ ( - \beta \ H \ ) \ ]_{av} \qquad (2)$$

Where $[.....]_{av}$ represents the average over all possible alloy configurations and $\beta = 1/kT$. The free energy can be expressed as

$$F = F_0 - kT \ [ \ ln < exp \ (- \beta \ V) > ]_{av} \qquad (3)$$

Where $F_0 = -kT \ [ln \ Tr \ exp \ (- \beta \ H_0 \ ) \ ]_{av}$ is the configuration averaged free energy for the system described by the Hamiltonian $H_0$. The Hamiltonian $H_0$ is so chosen that its eigenvalues are known. The symbol $<...>$ represents the ensemble average over the states of $H_0$ and the operator $V = H - H_0$. The simplest assumption for $H_0$ is to take

$$H_0 = \sum_i I_i \ S_{iz} \qquad (4)$$

The approximation then involves in evaluating the second term of eq.(3). It is assumed that the system is built out of small " building block" and use of $I_i$ as variational parameter. The variational form of the free energy can be expressed as [ 24,14]

$$F_v = F_0 - kT \sum_{n=1}^{L} \ [ln < ex(-\beta V_n) >]_{av} \cdots\cdots\cdots (5)$$

Where $V_n$ is the $n$-th division of $V$ which is divided into L number of blocks. With decrease of number (L) of division, $F_v$ tends to exact value of free energy. With the increase in size of the block the algebraic complexity grows at faster rate compared to improvement of the result related to transition temperature of an Ising model. Here we take the building block consisting of 4-spins for the disordered mixed spin system. The possible atomic configurations for 4-spin block with A or B distributed at lattice points are shown in Fig.-1. The respective probability of occurrence of the configuration is noted below the respective figure .

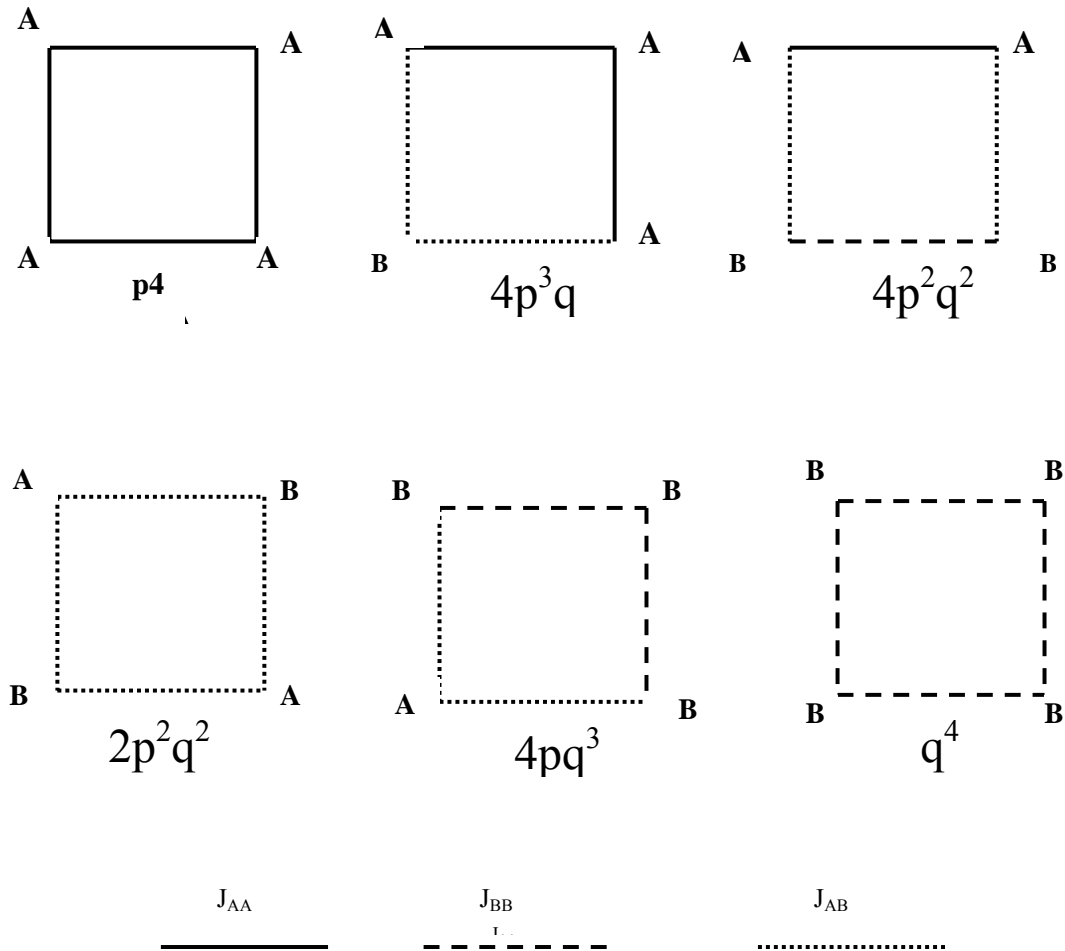

Fig.-1. Schematic representation of 'building block" of four atoms. Sold, Dashed and Dotted lines represent respectively $J_{AA}$,$J_{BB}$ and $J_{AB}$. The probabilities of different configurations are given below the diagrams.

The number of spin states in a given configuration depends on number of A and B atoms. The number varies from maximum ($3^4$) for block with all S =1 atoms to minimum ($2^4$) for all A-atom block. In this approximation the configuration - averaged trial free energy $F_v$ can be expressed as

$$F_v = F_0 - (z/8\beta) [ p^4 \ln Z_0 + q^4 \ln Z_4 + 4 p^3 q \ln Z_1 + 4pq^3 \ln Z_3$$
$$+ 2 p^2 q^2 \{ \ln Z_2 + 2 \ln Z_{22}\} ] \qquad .......(6)$$

Where

$$F_0 = - ((1-z/2) / \beta) [ p \ln (\cosh (\beta I_A/2)) + q \ln (\cosh (\beta I_B/2)) ]$$

$$Z_0 = 2 [ X_A^4 \cosh 2\alpha_1 + 4 \cosh \alpha_1 + 2 + X_A^{-4} ]$$

$$Z_4 = 2 X_B^4 \cosh (4\alpha_2) + 8 X_B^2 \cosh (3\alpha_2) + 4 (3 + 2 X_B) \cosh(2\alpha_2)$$
$$+ 8 (3 + X_B^{-2}) \cosh(\alpha_2) + 8 X_{BB}^{-1} + 9 + 2 X_B^{-4}$$

$$Z_1 = 2 X_A^2 X_C \cosh(\alpha_2 + 3\alpha_1 /2) + 2 X_A^2 X_C^{-1} \cosh( -\alpha_2 + 3\alpha_1 /2)$$

$$+ 2 [2 + X_A^{-2} X_C ] \cosh(\alpha_2 + \alpha_1 /2) + 2 [2 + X_A^{-2} X_C^{-1} ] \cosh( -\alpha_2 + \alpha_1 /2)$$

$$+ 2 [2 + X_A^{-2} ] \cosh(\alpha_1 /2) + 2 X_A^2 \cosh(3\alpha_1 /2)$$

$$Z_3 = 2 X_B^2 X_C \cosh(3\alpha_2 + \alpha_1 /2) + 2 X_B^2 X_C^{-1} \cosh( 3\alpha_2 - \alpha_1 /2)$$

$$+ 2[3 + X_B^{-2} X_C + 2 X_C^{1/2}] \cosh(\alpha_2 + \alpha_1 /2) + 2[3 + X_B^{-2} X_C^{-1} + 2 X_C^{1/2}] \cosh(\alpha_2 - \alpha_1 /2)$$

$+2[2X_B X_C^{1/2} + X_C]\cosh(2\alpha_2 + \alpha_1/2) + 2[2X_B X_C^{-1/2} + X_C^{-1}]\cosh(2\alpha_2 - \alpha_1/2)$

$+2[3 + 2X_B^{-1}(X_C^{-1/2} + X_C^{1/2}]\cosh(\alpha_1/2)$

$Z_2 = 2 X_C^2 \cosh(2\alpha_2 + \alpha_1) + 2 X_C^{-2}\cosh(2\alpha_2 - \alpha_1)$

$+4\cosh(2\alpha_2) + 6\cosh(\alpha_1) + 8\cosh(\alpha_2) + 6$

$+4 X_C \cosh(\alpha_2 + \alpha_1) + 4 X_C^{-1}\cosh(\alpha_2 - \alpha_1)$

$Z_{22} = 2 X_A X_B X_C \cosh(2\alpha_2 + \alpha_1) + 2 X_A X_B X_C^{-1}\cosh(2\alpha_2 - \alpha_1)$

$+4X_B X_A^{-1}\cosh(2\alpha_2) + 2 X_A[1 + 2X_B^{-1}]\cosh(\alpha_1)$

$4 X_A X_C^{1/2}\cosh(\alpha_2 + \alpha_1) + 4 X_A X_C^{-1/2}\cosh(\alpha_2 - \alpha_1)$

$+2[2X_A^{-1}(X_C^{-1/2} + X_C^{1/2}]\cosh(\alpha_2) + 2X_A^{-1} + 2 X_A^{-1}X_B^{-1}(X_C^{-1} + X_C)$

$\alpha_1 = [I_A(1-2z^{-1}) + 2 H_A/z]\beta$ , $\alpha_2 = [I_B(1-2z^{-1}) + 2 H_B/z]\beta$ , z = number of nearest neighbours,

$X_A = \exp(\beta J_{AA}/4)$ , $X_{BB} = \exp(\beta J_{BB})$ and $X_C = \exp(\beta J_{AB})$.

In above expressions of $\alpha$'s symbols H's and I's represent respectively the magnetic field and the variational parameters at respective site.

The variation fields are determined from the minimization of the trial free energy $F_v$ with respect to variational parameters $I_A$ and $I_B$ :

$$\delta F_v / \delta\, I_A = 0 \qquad \text{and} \qquad \delta\, F_v /\delta\, I_B = 0$$

This leads to the coupled equations for $I_A$ and $I_B$ as

$$2\tanh(\beta I_A/2) = p^3(A_0/Z_0) + 4p^2q(A_1/Z_1) + 4q^3(A_3/Z_3)$$

$$+2\,pq^2\{(A_2/Z_2) + 2(A_{22}/Z_{22})\} \qquad (7)$$

$$4[2\sinh(\beta I_B/2)]/[1+2\cosh(\beta I_B/2)] = q^3(B_4/Z_4) + 4p^3(B_1/Z_1) + 4pq^2(B_3/Z_3)$$

$$+ 2p^2q\{(B_2/Z_2) + 2(B_{22}/Z_{22})\} \qquad (8)$$

where the expressions for A's and B's are listed in Appendix-1.

The self-consistent solutions of the equs. (7 −8) provide non-trivial values of I's ,and the approximate free energy is then found in substituting the solutions in the expression of $F_v$(equ.6). The sub-network magnetization per atom then comes out as

$$M_A = 0.5 \tanh(\beta\, I_{A0} / 2) \qquad (9)$$

$$M_B = 2 \sinh(\beta\, I_{B0} /2) /[1 + 2\cosh(\beta\, I_{B0} /2)] \qquad (10)$$

and the total magnetization per atom $M = p\, M_A + q\, M_B$ . (11)

Here $I_{A0}$ and $I_{B0}$ are the self-consistent solution of the coupled equations (7) and (8). The finite values of $I_{A0}$ and $I_{B0}$ lead to spontaneous sub-network magnetization which appears below the transition temperature $T_c$. The equation for the transition temperature follow from the condition that the sub-network magnetization approaches to zero as T tends to $T_c$. The equation for $T_c$ that follows from eqs.(7 -10) is lengthy ,but in the limit of large z the transition temperature $T_c$ comes out as

$$T_{cM} = \{ p\,T_A + (1-p)\,T_B + [(\,p\,T_A - (1-p)\,T_B\,)^2 + 4p(1-p)\,T_C{}^2]^{1/2} \}$$

(11)

with  $T_A = J_{AA} S_A(S_A+1)\, z\,/\,6k$  ,  $T_B = J_{BB} S_B(S_B+1)\, z\,/\,6k$

$T_C = J_{AB}\ \{S_A(S_A+1)\, S_B(S_B+1)\}^{1/2}\ z\,/\,6k$

The transition temperature $T_{cM}$ is the mean field value for ferrimagnet with sublattice spin $S_A$ and $S_B$ [25]. In the following ,we present the numerical results taking z =8 and other model parameters like kT, magnetic field and the exchange interactions are expressed in terms of largest among three exchange parameters.

**RESULTS AND DISCUSSIONS**

**a) Magnetization :**

The ferrimagnetic systems is characterized by the thermal behaviour of spontaneous magnetization M and different types of M-T behaviour result depending on relative strength of the exchange interactions of the system. In this work we consider intra-sub-network interactions are of ferromagnetic nature ($J_{AA}$ and $J_{BB} > 0$) whereas inter-sub-network interaction antiferromagnetic ($J_{AB} < 0$). The effect of the antiferromagnetic interaction on the magnetic behaviour would be maximum for system with p=0.5 and we treat this case first with different sets of exchange interactions. At outset the system with $J_{AA} > |J_{AB}| > J_{BB}$ is considered. This mimics the rare–earth –transition metal series where the ferromagnetic exchange interaction between 3d atoms is large compared to that between rare-earths. Moreover, the exchange interaction between rare-earth and 3d–atom is weakly antiferromagnetic. In Fig.2 we present the thermal behaviour of sub-network magnetization ($M_A$ and $M_B$) and total magnetization M per magnetic atom for $J_{BB} = 0.02$ and $J_{AB} = - 0.1$ ($J_{AA} =1$). At low T, the respective sub-network magnetization saturates to its full value. As T increases $M_A$ diminishes smoothly whereas $M_B$ changes abruptly to a small value at a temperature T $\cong$ 0.15. This results a compensation temperature $T_{cm}$ where the total magnetization M suffers a sign reversal. Thermal variation of $M_B$ for $T_{cm} < T < T_c$ is unusual and can not be fitted with the characteristic thermal behavior of Ising ferromagnet. Finally, the sub-network magnetization vanishes at ferrimagnetic transition Tc

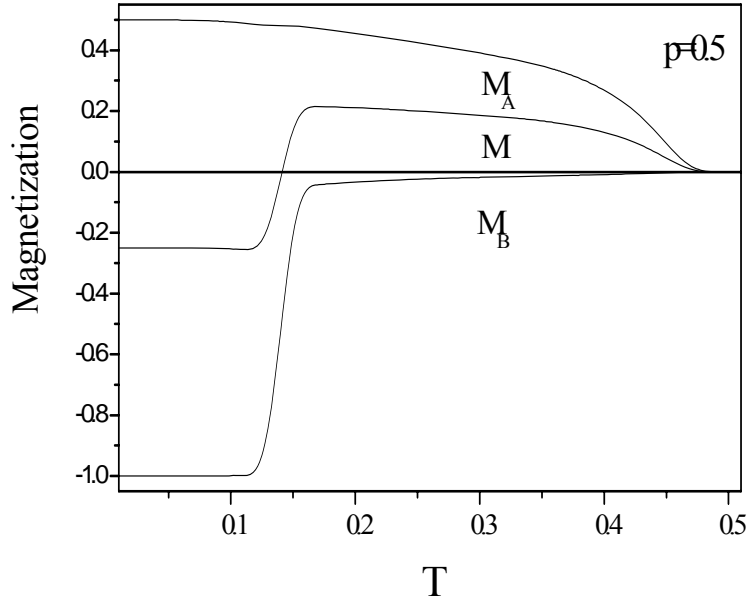

Fig-2. Variation of sub-network magnetization $M_A$ and $M_B$ and total magnetization M with temperature T ( T $\cong$ kT/$J_{AA}$) for p=0.5 , $J_{AB} = - 0.1$, $J_{BB} = 0.02$ and $J_{AA} =1.0$ (Unless mentioned
$J_{AA} =1.0$ is taken in following figs.)

This is typical of N-type ferrimagnet [26] .The ferrimagnetic alignment of magnetization is due to combining effect of strong ferromagnetic $J_{AA}$ and antiferromagnetic $J_{AB}$. The switching-like characteristic of $M_B$ (and M) is due to small value of $J_{BB}$ . Due to weak $J_{BB}$, the B-sub-network becomes nearly demagnetized when the magnetic energy as determined by $J_{BB}$ is comparable to thermal energy kT.

The effect of inter-sub-network interaction on behaviour of the magnetization (M) is depicted in Fig.-3 for different values of $J_{AB}$ ranging from -0.01 to -0.4.

The perfect anti-parallel spin alignment between A- and B-networks persists up to a certain temperature. This temperature interval increases as two networks are coupled more strongly. With the increase in magnitude of $J_{AB}$ the sharpness of the magnetization reversal at $T_{cm}$ goes down and the net magnetization M for T lies between $T_{cm}$ and $T_c$ becomes smaller. Above a critical value of $J_{AB}$ that depends on values of other exchange constants there is no compensation regime and for chosen set of J's it is $-0.3$. The magnetization reversal occurs when spin fluctuation in sub-network of higher moment is stronger compared to that in other one.

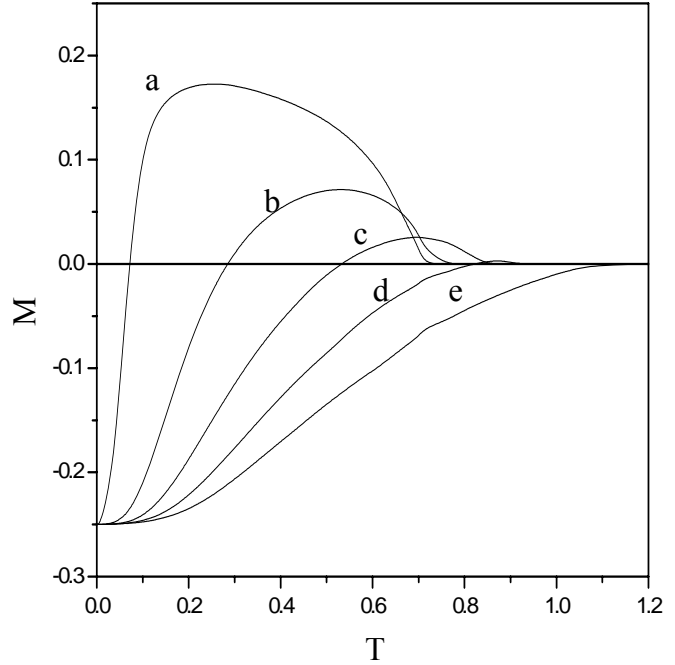

Fig.-3. Total magnetization M vs. temperature T for p=0.5, $J_{BB} = 0.02$ and different values of $J_{AB}$ = -0.01 (a), - 0.1 (b),- 0.2 (c),-0.3 (d),-0.4 (e)

Next we examine the system where $J_{AA}$ and $J_{BB}$ are ferromagnetic in nature and comparable in magnitude ($J_{AA}$ =$J_{BB}$ =1) with different values of antiferro-$J_{AB}$. The thermal variation of spontaneous magnetization is depicted in Fig.4 for $J_{AB}$ = - 0.2 ( solid) and -1.0 (dashed) in terms of reduced temperature t = T/ $T_c$. For cases $J_{AB} \leq$ -1.0(dashed line), thermal variation of $M_A$, $M_B$, and M is similar to ferromagnetic system and monotonically vanishes at T = $T_c$ and refers to Q category of ferrimagnetism. For $|J_{AB}| < J_{AA}$ , sharper decreases in $M_A$ above $t \approx$ 0.3 is found ,whereas $M_B$ exhibits normal variation. Due to this, M passes through broad maximum. The maximum value of M becomes larger and position of the maximum shifts to lower temperature for smaller value of $J_{AB}$. This corresponds to P-type ferrimagnet [26]. Despite same strength of exchange interaction the internal magnetic energy in sub-network A is less than that in sub-network B due to higher spin value of $S_B$ and this difference in energy leads to higher fluctuation in $M_A$ at finite temperature. The maximum in M is the consequence of combine effect of larger thermal fluctuation in $M_A$ and higher value of $M_B$. This effect is more pronounced when $J_{AA} << J_{BB}$ and a representative results with $J_{AA} = 0.2$ and $|J_{AB}| \leq J_{BB} = 1.0$ is shown in Fig. 5.

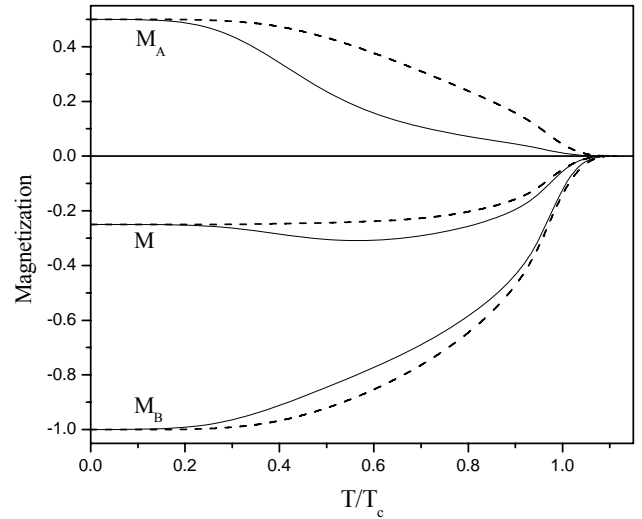

Fig.4-Variation of sub-network magnetization $M_A$ and $M_B$ and total magnetization M with reduced temperature T/$T_c$ for p=0.5 with $J_{BB}$ = $J_{AA}$ =1.0 and $J_{AB}$ = - 0.2(solid) −1.0 (dashed).

When A-B coupling is weak (solid curve) , A-sub-network is nearly disordered for T << $T_c$ and system shows higher magnetization when heated from low temperature . However, with increase in A-B coupling the spin fluctuations in A-sub-network is inhibited (dotted curve) and $M_A$ remains finite upto t =1 and M monotonically decreases with t. This is the consequence of higher magnetic energy available from antiparallel arrangement of magnetizations due to stronger network coupling.

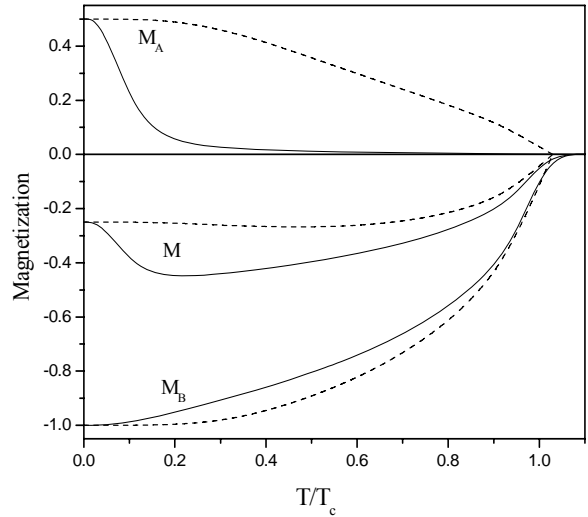

Fig.-5.Variation of sub-network magnetization $M_A$ and $M_B$ and total magnetization M with reduced temperature $T/T_c$ for p=0.5 with $J_{BB} = 1.0$ ,$J_{AA} = 0.2$ and $J_{AB}$ = - 0.1(solid) −1.0 (dashed).

The compositional dependence of the magnetic behaviour is examined by varying p with fixed $J_{AA} = 1$,$J_{AB}$ = - 0.1 and $J_{BB} = 0.02$ and the results are depicted in Fig.-6.At low temperature the spins within the sub-network are ferromagnetically coupled and those of different sub-networks are antiferromagnetically aligned.

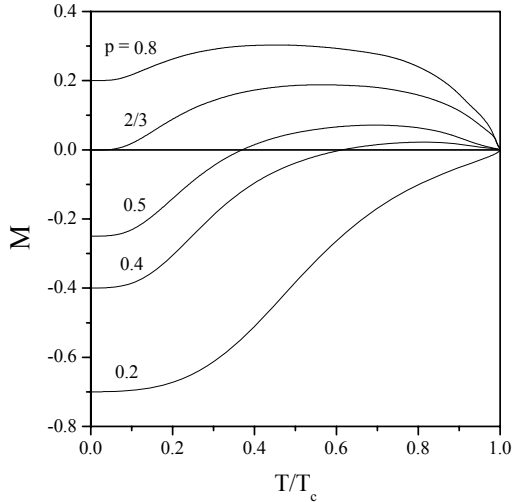

Fig.-6. Variation of magnetization M with reduced temperature T/T$_c$ for different concentration p with J$_{BB}$ = 0.02, and J$_{AB}$ = - 0.1.

For T /T$_c$ ≤ 0.16 (this value is determined by J$_{BB}$ and z) the total magnetization M saturates to (-1 + 3p/2) , and thermal behaviour of M depends on concentration. For 0.2< p < 2/3, M changes its direction at compensation temperature T$_{cm}$ that decreases with decrease in p. Above T$_{cm}$ the sub-network-A preserves higher magnetization due to stronger intra exchange interaction. In contrast , the spin fluctuation in the sub-network-B caused by smaller value of B-B exchange reduces M$_B$ and is the reason for a broad maximum of M between T$_{cm}$ and T$_c$. At higher concentration the system becomes more magnetized at intermediate temperature as B network is thermally demagnetized in faster rate. This is exemplified in Fig.7a for A-rich alloy (p = 0.8) for J$_{BB}$= 0.2 and J$_{AB}$= - 0.02 (solid) and = -1.0(dashed). While thermal behaviour of M$_A$ remains nearly same, that of M$_B$ becomes very different for two cases. It turns out from these results that M exhibits a broad maximum for p-rich system when one intra-sub-network exchange is stronger compared to other exchange interactions. For B-rich region (p=0.2) and J$_{AA}$ small compared to J$_{BB}$ the role of M$_A$ and M$_B$ is interchanged (Fig. 7b).

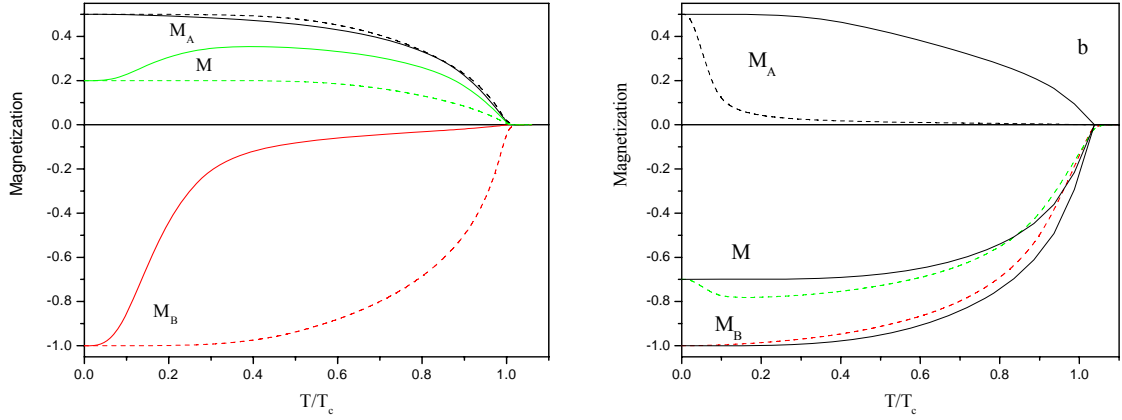

Fig.7 Plot of sub-network magnetization $M_A$ and $M_B$ and total magnetization M as function of reduced temperature $T/T_c$ : a) for p=0.8 and $J_{BB}$ = 0.2 with $J_{AB}$ = -1.0 (dashed) and –0.02 (solid) $J_{AA}$ =1.0. b) for p=0.2 and $J_{AB}$= -0.5.Solid curve is with $J_{AA}$ =1.0 and $J_{BB}$ = 0.2 , and dashed for $J_{AA}$ =0.2 and $J_{BB}$ = 1.0.

.However, despite smallness of A-A interaction $M_A$ exhibits normal thermal depaendence for stronger A-B interaction. In this situation magnetic energy associated with antiparallel orientation of magnetizations dominate

A special situation arises for p=2/3 in mixed $S_A$=1/2 and $S_B$=1 system .For this composition of the alloy, the magnetization is fully compensated at low temperature for all vales of exchange parmeters. On the other hand the net magnetization at intermediate temperature depends on relative strength of inter – and intra – sub-network interaction. For strong $J_{AB} \leq$ -0.75 ($J_{AA}$ =1. and $J_{BB}$ =0.5) the compensation (M=0) situation persists over wide range of $T/T_c$ (Fig.8-dashed curve). Close to $T_c$ , there is deviation from complete compensation and small magnetization appears with $M_B > M_A$ . This is similar to situation like antiferromagnet. For weak coupling ($J_{AB}$ = - 0.1) between sub-network and M remains essentially zero as T→0 . At intermediate temperature (T< $T_c$) fully compensated magnetic state breaks down and the magnetization becomes finite and dominated by sub-network magnetization $M_A$ due to higher value of $J_{AA.}$ This situation corresponds to transition from antiferro- to ferri- magnetic state.

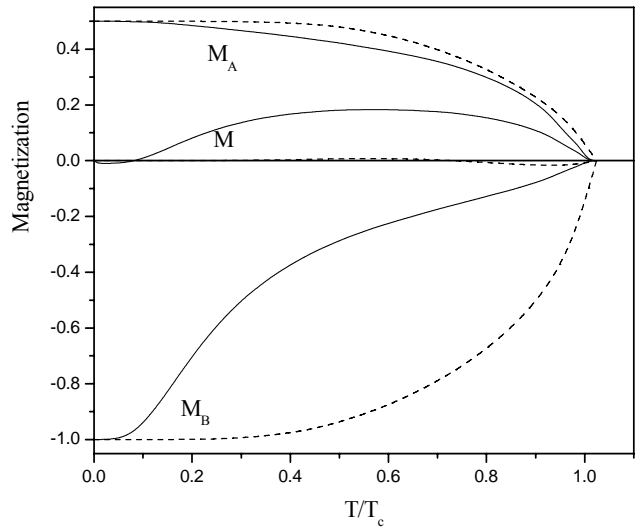

Fig.8. Thermal behaviour of magnetization $M_A$ and $M_B$ and magnetization M with reduced temperature $T/T_c$ for p= 2/3 with $J_{AA}$ =1.0 with $J_{BB}$ =0.5 , $J_{AB}$ = - 0.75 (dashed ) and $J_{BB}$ =0.02 , $J_{AB}$ =-0.1(solid)

### b) Transition temperature:

The para to ferri -magnetic transition temperature $T_c$ and the compensation temperature $T_{cm}$ are numerically obtained from M vs T result. The representative result is presented for p=0.5 considering intra-sub-network interaction $J_{AA}$ is higher compared to other two exchange interaction and only $J_{AB}$ is anti-ferromagnetic like ( Fig.-9a). In the plot all parameters are normalized in terms of $J_{AA}$ =1. The variation of $T_c$ with $J_{AB}$ ( $J_{BB}$ ) depends on the relative magnitude of $J_{BB}$ ( $J_{AB}$ ).

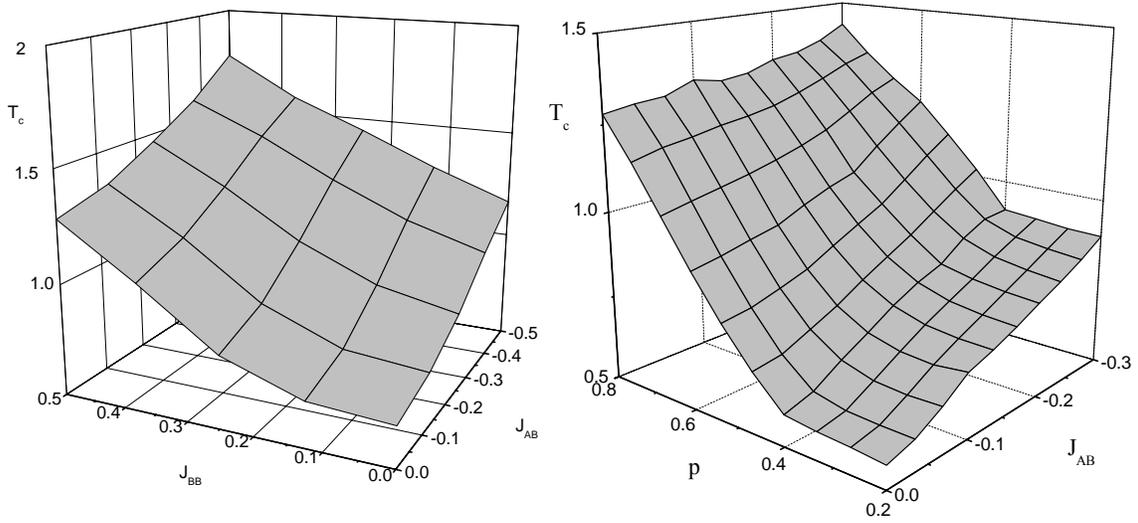

Fig. 9. Transition temperature $T_c$: a) for p=0.5 as functions of $J_{AB}$ and $J_{BB}$ : b) for fixed $J_{BB}$ =0.2 as functions of $J_{AB}$ and p.

When inter-sub-network interaction is weak, the variation of $T_c$ with $J_{BB}$ is highly non-linear. Non-linear increase in $T_c$ with $J_{AB}$ is also found for case of small $J_{BB}$. With the increase in magnitude of $J_{AB}$ non-linearity in variation of $T_c$ goes down for fixed $J_{BB}$. When $J_{BB}$ and $J_{AB}$ are comparable slower variation of $T_c$ is obtained. In Fig.9b the results of $T_c$ ( in unit of $J_{AA}$=1) for fixed value of $J_{BB}$ = 0.2 for different values of $J_{AB}$ and p are shown. The decrease in $T_c$ for alloy with smaller $J_{AB}$ is rapid with initial decrease of concentration but for p < 0.5 concentration dependence of $T_c$ becomes weak. Initial larger dependence is related to occurrence of more number of antiferro- bonds as p decreases , and for small p, $T_c$ becomes small due to presence of larger number of smaller B-B ferro-bonds. Within A-rich region weak dependence of $T_c$ on $J_{AB}$ is related small number of A-B and B-B magnetic bonds. Although $T_c$ is an increasing function of $J_{AB}$ for all concentration ,a stronger variation for B-rich alloy is due to smaller value of $J_{BB}$. Above results are the situation where intra-sub-network interaction ,namely $J_{AA}$ is dominant. Different concentration dependence is found when exchange interaction between A and B is larger than others and this situation corresponds to usual ferromagnetic system. Fig.10 depicts the concentration variation of $T_c$ (ferri- to para) for different combination of $J_{AA}$ and $J_{BB}$ (all normalized with $|J_{AB}|$ ). A broad maximum appears for $J_{AA}$ and $J_{BB}$< $|J_{AB}|$. For given $J_{BB}$, as $J_{AA}$ increases the position of maximum shifts towards higher p and also $T_c$ at maximum increases. In contrast the maximum is pushed towards lower p as $J_{BB}$ becomes higher. For $J_{BB}$ = $J_{AA}$ = 0.2 the maximum appears at p which is slightly less than 0.5 and this is related to the dependence of $T_c$ on spin and $S_B > S_A$. The shift of the maximum depends on relative number and magnitude of antiferro- and ferro- magnetic bonds. The hgher value of $T_c$ for p< 0.5 is again related to higher spin value of B – atom. It is to be noted that a maximum has been observed in amorphous Gd -Fe based alloy [8].

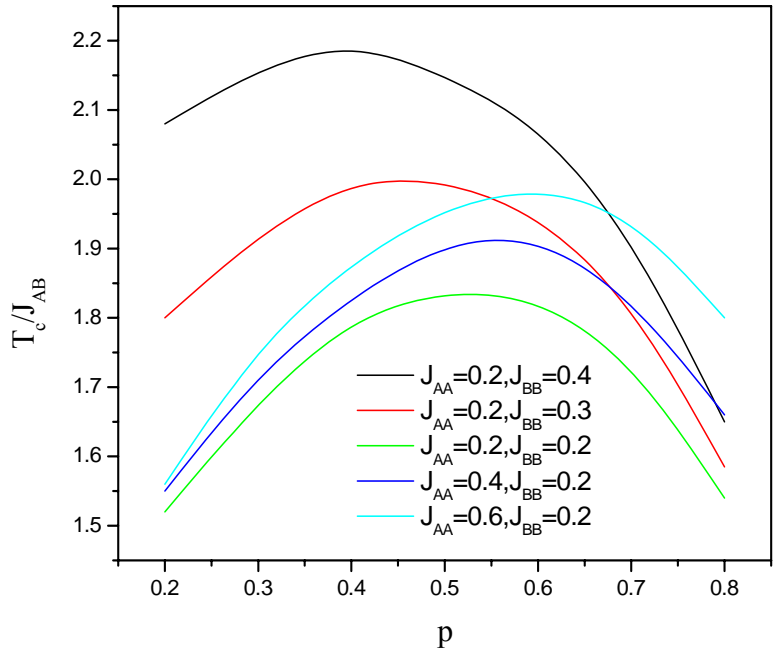

Fig. 10: Plot of $T_c$ as a function of concentration p for different
values of $J_{AA}$ and $J_{BB}$ with $J_{AB}$ = -1

### c )   Compensation temperature

The compensation temperature $T_{cm}$ is determined from the condition that M reverses its sign at $T_{cm}$ ($<T_c$). The phase boundary for $T_{cm}$ in $J_{AB}$ - $J_{BB}$ plane for p = 0.5 is shown in Fig.11a. The compensation temperature exists for alloy (p=0.5) with exchange parameters lying below the boundary line. In order to have finite $T_{cm}$ the B-sub lattice magnetization must decrease in faster rate and that would occur when both A-B and B-B magnetic bonds are weak. The variation of $T_{cm}$ with $J_{AB}$ and $J_{BB}$ for p = 0.5 is presented in Fig.11b. For given $J_{AB}$ , $T_{cm}$ is shifted towards a higher value as $J_{BB}$ increases. Similar trend of $T_{cm}$ is evident for fixed $J_{BB}$ and increasing $J_{AB}$ It follows from the numerical calculation that the compensation point exists for a small regime of $J_{AB}$ . The interval of $J_{AB}$ over which compensation point is found, shrinks with increase in $J_{BB}$ . The compensation point appears at lower temperature for p=0.6 compared p=0.5

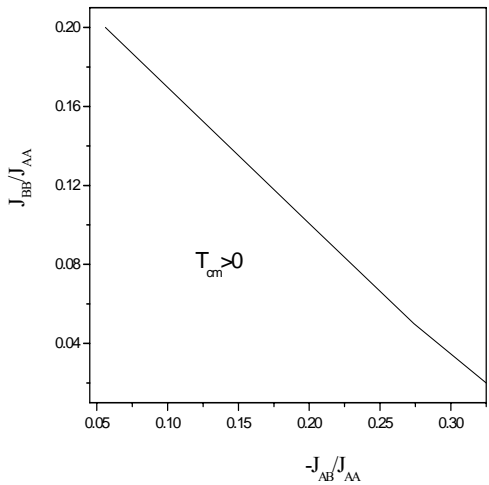

Fig.11a:Region in $J_{BB}$ –$J_{AB}$ plane for p=0.5 with $J_{AA}$ =1.0.The region below the line has finite compensation temperature

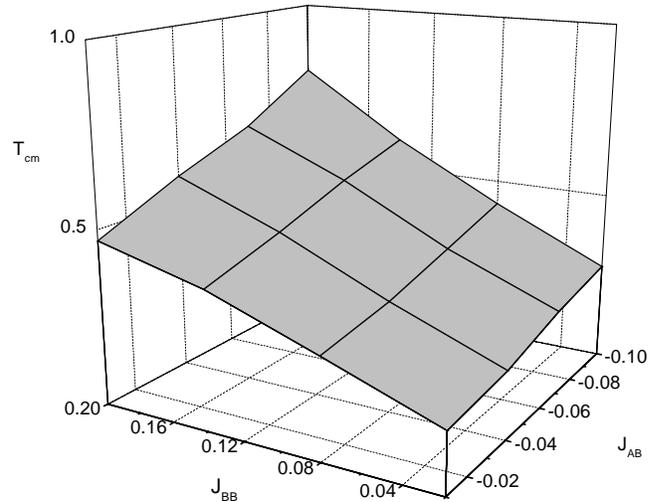

Fig.11b. Compensation temperature $T_{cm}$ for p=0.5 as functions of $J_{AB}$ and $J_{BB}$ ( $J_{AA}$ =1)

for same set of exchange parameters. For p > 2/3 and also smaller value of p there is no compensation temperature

Fig.12 compares the  transition temperatures $T_c$  and  $T_{cm}$  for p = 0.5   for two sets of   small values of $J_{BB}$ .  It is evident that the effect of variation of $J_{BB}$  on  $T_{cm}$  and $T_c$ is very different. The   nature of variation of $T_c$ with $J_{AB}$ is similar for both  values of $J_{BB}$  except that $T_c$  is slightly reduced  when $J_{BB}$ is reduced by a factor of ten. In contrast, dependency of $T_{cm}$ on $J_{AB}$ changes largely  for  above change of $J_{BB}$.For $J_{BB}$ = 0.2 , $T_{cm}$  exists only for limited range of  $J_{AB}$.

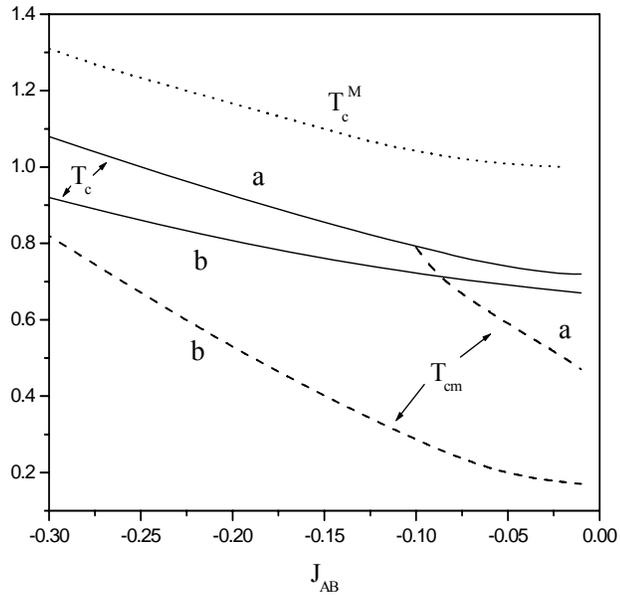

Fig.12.Variation of $T_c$ and $T_{cm}$  for p=0.5 with  $J_{AB}$  a) $J_{BB}$ =0.2.,b) $J_{BB}$ =0.02., $T_c^M$  is in  mean field approximation (equ.12) $J_{AA}$ =1.0,

For very small $J_{BB} = 0.02$ the compensation point persists over a broad range of $J_{AB}$ and difference between the compensation point and transition temperature goes down. We note that the magnetization reversal is abrupt when $T_{cm}$ is low. The transition temperature $T_c$ (normalized with $J_{AA} = 1.$) is compared with that due to the mean field approximation. The dotted curve $T_c^M$ is the mean field result of transition temperature for $p=0.5$, $J_{BB}=0.2$. The transition temperature in cluster variational method is much reduced compared to that of mean field case. The extent of reduction of $T_c$ depends on size of cluster and increase as the size of "building block" increases. The complexity of the calculation also grows with increase in size due to increase in number of spin states. The reduction in $T_c$ is only few percent when cluster of eight atom is considered [27].

## d) Magnetic susceptibility

The magnetic susceptibility $\chi$ is obtained numerically by calculating M at T> $T_c$ in the limit of small magnetic field H. The results for alloy with p = 0.5 where the exchange interaction $J_{AA}$ (=1) dominates (Fig.12). The calculated values of inverse of $\chi$ with $J_{BB}$= 0.02 and $J_{AB}$ = - 0.3 (a) and − 0.2 (b) is plotted as a function of T/$T_c$ (points). At high temperature (T/$T_c$ > 1) the susceptibility follows the Curie-Wiess behaviour , $\chi^{-1} \sim$ (T + θ ) where θ (>0) depends on exchange parameters. The divergence of susceptibility appears very close to $T_c$. It is interesting to note that the susceptibility for whole range of temperature can be fitted with a formula :

$$\chi^{-1} = \chi_0^{-1} + \frac{T}{C} - \frac{\Delta}{T - \Theta} \qquad (13)$$

where  parameters $\chi_0$ C ,$\Delta$, and $\Theta$ are independent of temperature but are function of p and exchange parameters. The dotted lines represent the plot of  the equ.(13) with $\Delta$ = 0 and fit very well at high temperature. The intercept  of the line determines $\chi_0$  and asymptotic  Curie temperature $T_c^A = -C \chi_0^{-1}$  is given by   $\chi_0^{-1} = 0$. The truncated equation  ($\Delta$ =0) fits the  calculated results  for T close to $T_c$ for system with stronger $J_{AB}$. The transition temperature  can be expressed in terms four  parameters of eq.(13) from the condition  $\chi^{-1} = 0$. In Fig.13, the corresponding result of magnetic susceptibility  of  ferrimagnetic system with $J_{AB}$ = -1.0 and $J_{AA}$=$J_{BB}$ = 0.2 is presented for p = 0.5. Here again calculated points are fitted well with equ.(13) however the contribution from last term remains finite for higher range of temperature. The structure of formula (13) is akin to the mean-field result for two-sublattice  ferrimagnet with identical moment [26] .

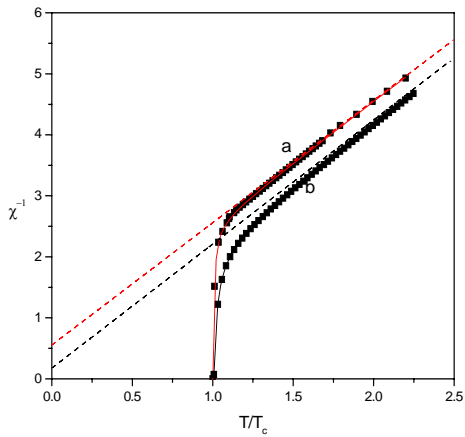

Fig.12. Plot of $\chi^{-1}$ in paramagnetic state as a function of reduced temperature T/$T_c$ for p=0.5 with $J_{BB}$ =0.02 :and $J_{AB}$ = - 0.3 (a), $J_{AB}$ =- 0.2(b)

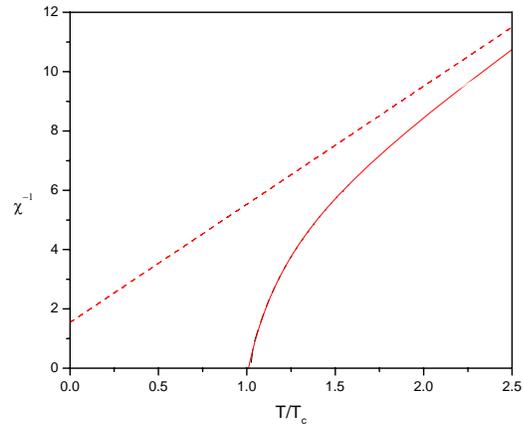

Fig.13 Plot of $\chi^{-1}$ in paramagnetic state vs T/$T_c$ for p=0.5 with $J_{AA}$ = $J_{BB}$ = 0.2 ,$J_{AB}$ = - 1.0.Symbols are calculated values, solid line fitted with formula (13) and dashed represents truncated equ.(13) with $\Delta$ = 0.

**e) Field Dependence of magnetization**

The  ferrimagnetic system  with uncompensated magnetization  exhibits  field dependence of M similar to ferromagnetic situation at field much smaller than the exchange field. At higher field the magnetization M is governed by competitive aspect of external and internal fields. .The result of M vs H  of alloy for p = 0.5 and $J_{BB}$ = 0.02 and $J_{AB}$ = - 0.1 and varying H from 0 to 0.5 in steps of 0.1 applied along +z-direction is depicted in Fig.14.

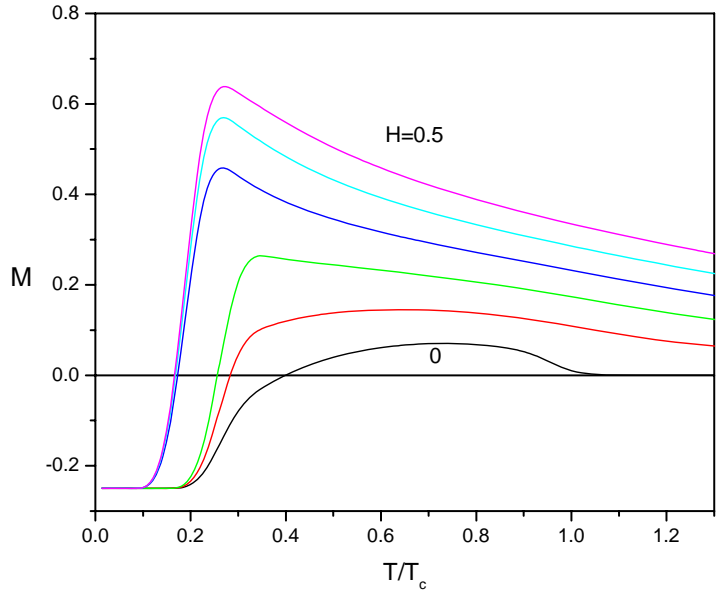

Fig.14 Field dependence of magnetization M for p=0.5 as function of $T/T_c$ $J_{BB} = 0.02$ , $J_{AB} = -0.1$. Field H  (H ≡H/$J_{AA}$) varies from 0 to 0.5 in  steps of 0.1.

As field magnitude increases  M deviates from its saturated  value  M = - 0.25 at low temperature. The temperature where magnetization reversal takes place shifts towards lower value of T. The  amount of shift is higher at lower field region and then varies very slowly at higher field.  Also at higher field region the magnetization  reversal  is steeper and a larger change of M at  compensation point appears. The magnetization  passes through a peak at a temperature that is decreasing function of field. The reversal  is related to energy balance between antiferromagnetically  coupled magnetizations  $M_A$ and $M_B$  and metamagnetic state of $M_A$ and $M_B$  created by field. At higher field metamagnetic state has lower  energy and larger value of M results from near parallel orientation of sub-sublattice magnetizations.   Thermal fluctuation has stronger demagnetization effect on  B-sub- sublattice  due to very weak exchange interaction and results M reversal. The magnetic field at first increases fluctuation in $M_B$  as field and $M_B$ are antiparallel but at higher field and higher temperature  higher magnetization is induced by magnetic field. More results on meta-magnetic behviour for other  cases will be reported in separate communication.

## 4.CONCLUSIONS

A disordered ferrimagnetic alloy ($A_pB_{1-p}$) with Ising like interaction between the spins ($S_A=1/2$ and $S_B=1$) of sub-lattices is treated using a cluster-variational method. In this method the interactions within the cluster of different configurations are treated exactly and the rest of the interaction is described by variational field which is obtained from minimization of free energy. The results on the magnetization, transition and compensation temperature and magnetic susceptibility are presented for different alloy and exchange parameters Different types of thermal behaviour of magnetization in ferrimagnetic state are obtained depending on relative strength of inter- and intra- exchange interaction. The transition temperature is much reduced compared to that follows from the mean field theory. The compensation temperature exists over a limited range of parameters. Both the compensation and transition temperature depend sensitively on inter-sublattice exchange and concentration The magnetic susceptibility above $T_c$ exhibits typical ferromagnetic behaviour and a formula is proposed for $T > T_c$. The meta-magnetic behaviour in presence of magnetic field is found ,and at high field the switching behaviour of magnetization with temperature is predicated. The alloy with strong exchange interaction within A-sub-lattice and weak one within B- and between sublattices closely represents the rare-earth-transtion metal alloy. like Fe-Gd alloy The result of such alloy shows compensation of magnetization for range of concentration p. For p> 0.5, the magnetization increases abruptly at low temperature.. Considering scanty experimental results and simple model for interaction it is expected that gap would exist between the observed and the calculated behaviour. However, we note that the results obtained here are in tune with general trend of the observation in this system. With increase in size of the "building block" the details of the thermal behaviour of magnetization and dependence of transition temperature on parameters would improve  the results  and will be reported later.

Acknowledgement: The author gratefully acknowledges help of I.Choudhuri   and Dr. P. Roychoudhuri for preparation of the manuscript

Appendix –1

$A_0 = 4 [ X_A{}^4 \sinh 2\alpha_1 + 2 \sinh \alpha_1 ]$

$A_1 = 3 X_A{}^2 X_C \sinh(\alpha_2 + 3\alpha_1 /2) + 3 X_A{}^2 X_C{}^{-1} \cosh( -\alpha_2 + 3\alpha_1 /2)$

$\quad + [2 + X_A{}^{-2} X_C ] \sinh(\alpha_2 + \alpha_1 /2) + [2 + X_A{}^{-2} X_C{}^{-1} ] \sinh( -\alpha_2 + \alpha_1 /2)$

$\quad + [2 + X_A{}^{-2} ] \sinh(\alpha_1 /2) + 3 X_A{}^2 \sinh(3\alpha_1 /2)$

$A_2 = 2 X_C{}^2 \sinh(2\alpha_2 + \alpha_1) - 2 X_C{}^{-2} \sinh(2\alpha_2 - \alpha_1)$

$\quad + 6 \sinh(\alpha_1) + 4 X_C \sinh(\alpha_2 + \alpha_1) - 4 X_C{}^{-1} \sinh( \alpha_2 - \alpha_1)$

$A_{22} = 2 \, X_A X_B X_C \, \sinh(2\alpha_2 + \alpha_1) - 2 \, X_A X_B X_C^{-1} \sinh(2\alpha_2 - \alpha_1)$

$\qquad + 4 \, X_A \, X_C^{1/2} \, \sinh(\alpha_2 + \alpha_1) - 4 \, X_A X_C^{-1/2} \sinh(\alpha_2 - \alpha_1)$

$\qquad + 2 \, X_A \, [1 + 2X_B^{-1}] \, \sinh(\alpha_1)$

$A_3 = \, X_B^2 X_C \, \sinh(3\alpha_2 + \alpha_1/2) - X_B^2 X_C^{-1} \sinh(3\alpha_2 - \alpha_1/2)$

$\qquad + [3 + X_B^{-2} X_C + 2 X_C^{1/2}] \sinh(\alpha_2 + \alpha_1/2) - [3 + X_B^{-2} X_C^{-1} + 2 X_C^{1/2}] \sinh(\alpha_2 - \alpha_1/2)$

$\qquad + [X_B \, X_C^{1/2} + X_C] \sinh(2\alpha_2 + \alpha_1/2) - [X_B X_C^{-1/2} + X_C^{-1}] \sinh(2\alpha_2 - \alpha_1/2)$

$\qquad + [3 + 2 X_B^{-1} (X_C^{-1/2} + X_C^{1/2}] \sinh(\alpha_1/2)$

$B_4 = 8 \, X_B^4 \, \sinh(4\alpha_2) + 24 \, X_B^2 \, \sinh(3\alpha_2) + 8 \, (3 + 2 \, X_B) \, \sinh(2\alpha_2)$

$\qquad + 8 \, (3 + X_B^{-2}) \, \sinh(\alpha_2)$

$B_1 = 2 \, X_A^2 X_C \, \sinh(\alpha_2 + 3\alpha_1/2) - 2 \, X_A^2 X_C^{-1} \cosh(-\alpha_2 + 3\alpha_1/2)$

$\qquad + 2 \, [2 + X_A^{-2} X_C] \, \sinh(\alpha_2 + \alpha_1/2) - 2 \, [2 + X_A^{-2} X_C^{-1}] \, \sinh(-\alpha_2 + \alpha_1/2)$

$B_3 = 6 \, X_B^2 X_C \, \sinh(3\alpha_2 + \alpha_1/2) + 6 \, X_B^2 X_C^{-1} \sinh(3\alpha_2 - \alpha_1/2)$

$\qquad + 2 [3 + X_B^{-2} X_C + 2 X_C^{1/2}] \, \sinh(\alpha_2 + \alpha_1/2)$

$\qquad + 2 [3 + X_B^{-2} X_C^{-1} + 2 X_C^{1/2}] \, \sinh(\alpha_2 - \alpha_1/2)$

$+ 4[X_B X_C^{1/2} + X_C] \sinh(2\alpha_2 + \alpha_1/2)$

$+ 4 [X_B X_C^{-1/2} + X_C^{-1}] \sinh(2\alpha_2 - \alpha_1/2)$

$B_2 = 4 X_C^2 \sinh(2\alpha_2 + \alpha_1) + 4 X_C^{-2} \sinh(2\alpha_2 - \alpha_1)$

$+ 8\sinh(2\alpha_2) + 8\sinh(\alpha_2) + 4 X_C \sinh(\alpha_2 + \alpha_1) + 4 X_C^{-1} \sinh(\alpha_2 - \alpha_1)$

$B_{22} = 4 X_A X_B X_C \sinh(2\alpha_2 + \alpha_1) + 4 X_A X_B X_C^{-1} \sinh(2\alpha_2 - \alpha_1)$

$+ 4 X_A X_C^{1/2} \sinh(\alpha_2 + \alpha_1) + 4 X_A X_C^{-1/2} \sinh(\alpha_2 - \alpha_1)$

$+ 8 X_B X_A^{-1} \sinh(2\alpha_2) + 2 [2X_A^{-1} (X_C^{-1/2} + X_C^{1/2}] \sinh(\alpha_2)$